\documentclass{article}

\usepackage{PRIMEarxiv}

\usepackage[utf8]{inputenc} 
\usepackage[T1]{fontenc}    
\usepackage{hyperref}       
\usepackage{url}            
\usepackage{booktabs}       
\usepackage{amsfonts}       
\usepackage{nicefrac}       
\usepackage{microtype}      
\usepackage{lipsum}
\usepackage{fancyhdr}       
\usepackage{graphicx}       
\usepackage{amsmath}
\graphicspath{{media/}}     

\title{Ratio1
\thanks{\textit{\underline{Citation}}: 
\textbf{Damian, A., Bleotiu, C., Butusina, P., De Franceschi, A., Toderian V., \& Grigoras, M. (2025). \textit{Ratio1}, \textit{Ratio1}.ai, DOI publication to be updated.}} 
}

\author{
  \begin{tabular}[t]{c}
    Andrei Damian, Cristian Bleotiu, Petrica Butusina \\
    Alessandro De Franceschi, Vitalii Toderian, Marius Grigoras \\\\
    \textit{Ratio1}.ai \\
    \texttt{\{author\}@ratio1.ai}
  \end{tabular}
}

\begin{document}
\maketitle

\begin{abstract}
We propose\footnote{Special acknowledgment and thanks to the whole Ratio1 team in section \ref{ACKS}} the \textit{Ratio1} AI meta-operating system (meta-OS), a decentralized MLOps protocol that unifies AI model development, deployment, and inference across heterogeneous edge devices. Its key innovation is an integrated blockchain-based framework that transforms idle computing resources (laptops, smartphones, cloud VMs) into a trustless global supercomputer \cite{Damian2025Ratio1}\cite{bleotiu2025naeuralaios}. The architecture includes novel components: a decentralized authentication layer (dAuth), an in-memory state database (\textit{CSTORE}), a distributed storage system (\textit{R1FS}), homomorphic encrypted federated learning (EDIL), decentralized container orchestration (Deeploy) and an oracle network (OracleSync) \cite{Damian2025Ratio1}\cite{bleotiu2025naeuralaios}, which collectively ensure secure, resilient execution of AI pipelines and other container based apps at scale. The protocol enforces a formal circular token-economic model combining \emph{Proof-of-Availability} (PoA) and \emph{Proof-of-AI} (PoAI) consensus: token rewards\footnote{Tokenomics and utilitarian token aspects in \ref{token1}, \ref{token11}, \ref{token111}, \ref{token112}, \ref{token113}, \ref{token12}} are computed by smart contracts as \begin{math}R(\omega_n) = f(\text{proof}(\omega_n))\end{math}, where each node’s uptime proof is validated on-chain \cite{Damian2025Ratio1} \cite{bleotiu2025naeuralaios}. Compared to centralized heterogeneous cloud MLOps and existing decentralized compute platforms, which often lack integrated AI toolchains \cite{Damian2025Ratio1} or trusted Ratio1 node operators (\textit{R1OP}) mechanics \cite{damian2025trust}, \textit{Ratio1}’s holistic design lowers barriers for AI deployment and improves cost-efficiency. We provide mathematical formulations of its secure licensing and reward protocols, and include descriptive information for the system architecture and protocol flow. We argue that our proposed fully functional ecosystem proposes and demonstrates significant improvements in accessibility, scalability, and security over existing alternatives.

\end{abstract}


\section{Introduction}
Pervasive computing envisions seamless integration of AI into everyday devices, yet current solutions rely on centralized clouds, leading to high costs and complexity. Modern AI applications promise intelligent services across industries, but widespread adoption is hindered by centralized infrastructure and high complexity. Contemporary cloud platforms (AWS, Azure, Google Cloud) provide robust GPU orchestration, yet are fundamentally centralized, subscription-based, and opaque.
These solutions incur prohibitive costs and require specialized DevOps expertise, placing AI capabilities largely in the hands of large providers. Edge and IoT devices remain largely underutilized for AI inference or training, creating a missed opportunity for scalability and cost reduction.

\subsection{Technical scope and reader advisory}
The majority of this whitepaper presents technical content designed for readers with  understanding of blockchain technology, cryptographic protocols, distributed systems architecture and machine learning operations. For utility token related aspects and overall utilitarian aspects of the circular tokenized economy promoted by Ratio1 please see sections \ref{token1}, \ref{token11}, \ref{token111}, \ref{token112}, \ref{token113}, \ref{token12}. These sections provide comprehensive documentation of the mining rewards vs fee distributions as well as native utility token creation, distribution protocols, and overall economic incentive structures. This approach ensures transparency and credibility while maintaining the rigorous technical standards

\subsection{The AI meta-OS approach}
Ratio1 builds upon prior decentralized ubiquitous computing research by introducing a meta-OS that combines blockchain transparency with edge compute orchestration.  The \textit{Ratio1} protocol’s utility-token model turns idle or partially used devices into income-generating assets, democratizing AI access for startups, developers, and underserved communities.  Key components include OracleSync for fault-tolerant monitoring, smart-contract based ND Licenses (ERC721), and R1 tokens (ERC20) for incentives\footnote{Through the dual incentive architecture: PoA mining emissions and PoAI fees} - followed by decentralized secure authentication (dAuth), decentralized encrypted file storage (\textit{R1FS}), heterogeneous federated learning based on homomorphic encryption (EDIL), in-memory-like key-value databases (\textit{ChainStore}) \cite{Damian2025Ratio1}.  The protocol abstracts infrastructure management via Infrastructure-as-Code and low-code SDKs, enabling rapid deployment of AI microservices through Ngrok-backed endpoints.  This whitepaper covers the technical and economic challenges addressed by \textit{Ratio1} and outlines the roadmap for a truly decentralized AI operating environment.

Existing work has proposed decentralized compute sharing via blockchain and fog computing. For example, in our previous published research we noted that tokenization and low-code workflows can automate large-scale ML pipelines while enabling peer-to-peer resource sharing. Nevertheless, most prior frameworks address only parts of the problem. General-purpose cloud systems emphasize scalability but lack peer-to-peer deployment or advanced privacy. Conversely, emerging decentralized platforms (e.g., iExec, Golem) match compute supply with demand via tokens, but focus on raw job execution and omit higher-level AI abstraction as well as trusted and compliant execution environments. Agent-based AI marketplaces (SingularityNET, Ocean) focus on models and data exchange rather than end-to-end deployment.

The \textit{Ratio1} Protocol aims to bridge these gaps by providing a \emph{decentralized AI meta-OS}: a unified, trustless platform that automates end-to-end MLOps on distributed hardware as well as traditional DevOps albeit in a decentralized heterogeneous orchestration approach. It integrates blockchain smart contracts, an edge-node software stack, and token incentives to overcome cost and integration challenges. In particular, \textit{Ratio1}'s contributions include: 
\begin{itemize}
    \item A novel licensing and reward mechanism combining PoA and PoAI to ensure genuine node participation and fair compensation via a pure utility token; \\
    \item A suite of decentralized infrastructure components (oracles, distributed storage, authentication, decentralized orchestration) that replace monolithic cloud services; \\
    \item Formal protocols for secure job orchestration and data privacy (e.g., encrypted data inference). 
\end{itemize}
We will describe these contributions in detail, comparing to related approaches, and support our discussion with precise mathematical models and reference designs.

\subsection{Related work }

Decentralized computing and blockchain-based AI have been explored in various contexts, but rarely as a complete MLOps solution. Major cloud AI services (AWS SageMaker, Google Cloud AI)\cite{hardt2021sagemaker} \cite{googlecloudai2025} provide scalable GPU/CPU resources on demand, but remain centralized, subscription-bound, censorship-oriented and with unclear data ownership policies. They lack native support for peer-to-peer deployment, lightweight plugin interfaces, or advanced privacy features (such as homomorphic encryption) beyond proprietary enclaves.

Several blockchain platforms (iExec\cite{fedak2018iexec}, Golem\cite{golem2024scientific}, Render Network\cite{rendernetwork2024}, Aethir\cite{aethir2025}, DeepBrain Chain\cite{deepbrainchain2017}, Theta\cite{theta2025}, Bittensor\cite{bittensor2024framework}) tokenize idle CPUs/GPUs to create decentralized compute markets \cite{damian2025trust}\cite{Damian2025Ratio1}. While these match resource providers with job requesters, they typically target raw compute orchestration. Most lack built-in microservice load balancing, AI-specific plugins, or no-code development tools or advanced tools such as decentralized in-memory key-value stores or secured decentralized storage. For example, iExec and Golem require users to package code manually and offer no native support for model templates or encrypted inference.

Decentralized AI and agent networks (e.g., Fetch.ai, SingularityNET, Ocean Protocol) introduce marketplaces for AI services, multi-agent coordination, and data exchange. These systems, however, focus on model sharing and agent autonomy, rather than providing a full-stack platform for data acquisition, model training, and deployment in a trustless or quasi-trustless environments. In practice, they often depend on specialized enclaves or custom hardware for privacy, limiting flexibility. Edge and fog computing frameworks (Akash, Aethir) enable decentralized container scheduling and on-premise device management at scale. While Akash offers a decentralized marketplace for general-purpose containers, it does not incorporate homomorphic inference or a microtransaction-based AI workflow out of the box. 

\subsection{Trusted vs trustless}
While AWS can (and will) suspend accounts for things like hosting illegal content, launching DDoS attacks, for known decentralized projects analyzed it is unclear what really happens with illegal actions - for example when users deploy critical applications with highly confidential data such as medical applications or fin-tech systems - once the harm was done \cite{damian2025trust}. 

Ratio1 introduces identity and governance from the start. In effect, \textit{Ratio1}’s compute providers are more like franchisees or licensed operators rather than random anonymous participants. This is a model we see in some “decentralized” industries: for instance, ride-sharing companies often require driver background checks and can deactivate drivers who don’t meet standards, even though drivers use their own cars. Similarly, \textbf{\textit{Ratio1} requires KYC for node operators (\textit{R1OP})} and can deactivate those who don’t meet the standards.

In summary, existing solutions either emphasize generic infrastructure or AI marketplaces. Few, if any, provide an integrated MLOps stack with edge data processing, smart contracts, and token economics. \textit{Ratio1} addresses this gap by combining the strengths of prior work into a holistic, decentralized AI operating system.

\section{Technical Framework and Architecture}

\subsection{Top view architecture}
Ratio1's architecture is structured as a layered meta-OS with three main architectural layers coordinating all operations:

\subsubsection*{\textit{Resource Layer}}
The Resource Layer consists of a distributed, heterogeneous network of Ratio1 Edge Nodes (REN) spanning consumer devices and servers. Each node runs a lightweight Ratio1 agent that advertises its compute capabilities (CPU, GPU, memory) and enables participation in a peer-to-peer cluster with no central coordinator. Nodes communicate via a proprietary protocol built on IoT standards (e.g., MQTT\cite{hunkeler2008mqtt}\cite{amqp}\cite{mqtt}) and peer-to-peer storage backends like IPFS, with all messages secured by digital signatures. This ensures that Ratio1 agents can trustlessly verify commands and data from peers, leveraging blockchain-grade immutability and signature validation for security. 

The hardware environment is highly diverse - from laptops to data-center GPUs - so the system enforces baseline requirements (64-bit CPUs, multi-core, 16GB+ RAM, 100GB+ local storage) to maintain performance uniformity. Importantly, any device with spare capacity can become an AI worker node, contributing to a global “compute fabric” that executes tasks closer to end-users (reducing latency) and taps into otherwise idle resources. Ratio1’s design exploits this heterogeneity for resilience and performance: nodes can join or leave dynamically, and loads are shifted transparently to available nodes, avoiding any single point of failure, albeit penalizing rewards for the \textit{R1OP}s that do not provide sufficient support for their nodes. 
Each Edge Node is cryptographically identified by an EVM-based key pair (with a corresponding on-chain license NFT), enabling secure node onboarding and governance. 

Overall, the resource layer transforms everyday hardware into a decentralized AI supercomputing grid, with Ratio1 agents orchestrating local container execution and interfacing with the network’s consensus protocols to maintain a trustless, high-availability infrastructure.

\subsubsection*{\textit{Service Layer}}

On top of the raw nodes, Ratio1 provides a Service Layer composed of specialized, decentralized services that coordinate computation, storage, security, and consensus across the network. These core services communicate through the Ratio1 protocol (via the P2P network and on-chain signals) to ensure reliable, fault-tolerant operations. 

Key components include:
\begin{itemize}
    \item Deeploy (Decentralized managed orchestration) - Deeploy is Ratio1’s container orchestration system, analogous to a cloud controller but without any central server. Instead of a Kubernetes master, Deeploy relies on smart contracts and a network of validator nodes to schedule tasks and handle lifecycle events;
    \item ChainDist (Distributed Scheduler) - ChainDist provides Ratio1 native distributed job scheduling engine. It replaces the centralized scheduler of traditional systems with a blockchain-powered scheduler.
    \item Ratio1 \textit{CSTORE} (\textit{ChainStore}) - \textit{CSTORE} is a decentralized in-memory key-value store, akin to a distributed Redis cache, which provides fast state sharing across the network without a central database. It is designed to prevent single points of failure in data caching and coordination by replicating and partitioning key-value pairs among participating nodes.
    \item \textit{R1FS} (Distributed File System) - \textit{R1FS} is Ratio1’s decentralized file storage service, built atop the IPFS protocol to leverage content-addressable, peer-to-peer storage. It provides a global file system in which files are identified by cryptographic hashes (CIDs) and stored redundantly across the network rather than in any single server. \textit{R1FS} is private to the Ratio1 network (data can be encrypted and shared only among permissioned nodes), but it inherits IPFS’s properties like deduplication, versioning, and hash-based integrity verification.
    \item dAuth (Decentralized Authentication) - dAuth is a global authentication and instance provisioning service for Ratio1, designed to securely onboard both Edge Nodes and client SDKs without traditional credential management.
    \item R1 OracleSync (Consensus \& Oracle Integration) - OracleSync is the consensus layer that connects off-chain events and on-chain trust, using a network of oracle nodes, supervisor-like nodes ran by various founders, team members and early adopters, to validate tasks and state in a trustless manner.
\end{itemize}

\subsubsection*{\textit{Application Layer}}
At the highest level, the Application Layer offers developers and end-users a platform to build AI workflows and services with minimal complexity. Ratio1’s meta-OS abstracts the distributed infrastructure behind a suite of low-code/no-code tools and SDKs. This empowers users to compose complex AI-driven applications (e.g., data processing pipelines, model inference services, IoT analytics) without needing to manage hardware, containers, or scale. One cornerstone of this layer is a low-code pipeline framework: applications can be assembled from modular plugins representing data sources, ML models, and business logic, configured via a graphical interface or simple JSON or YAML specifications.

Internally, the system breaks down an AI application into distinct subsystems - communication, data acquisition, model serving, and business process - each of which can be fulfilled by pre-built plugin components. For instance, a user could select an image feed plugin (for camera input), connect it to a pre-trained object detection model plugin, then to a notification plugin that sends results to a dashboard. Ratio1’s platform handles the wiring of these components, so data flows from one to the next in a distributed execution pipeline. Thanks to this design, altering an application’s behavior (say, swapping a model or changing a data source) is as easy as changing the plugin configuration, with no need to write integration code. This dramatically simplifies MLOps for multi-node deployments - the heavy lifting of parallelizing tasks, handling message queues, and synchronizing results is managed by the platform. In addition to the visual/low-code builder, Ratio1 provides SDKs (currently in Python and Node.js, among others) for developers who prefer to script their workflows.

The Ratio1 SDK, allows from launching containers to executing distributed functions with just a few lines of code as it abstracts away the details of contacting dAuth, packaging code into a container, and communicating with Edge Nodes. A developer can call high-level methods like \verb|session.create_container_app(...)| and \verb|app.deploy()| to launch a microservice, and the SDK will automatically handle authentication, job submission to Deeploy, and even log retrieval. This means AI practitioners can use familiar programming workflows to deploy to Ratio1’s network just as they would to a local Docker or cloud instance, without learning new paradigms. 

Finally, the Application Layer also includes ready-made AI services, agents and even end-to-end applications that developers can incorporate, fork and pivot. These come as off-the-shelf full applications templates, microservices or just plugins that can be forked or even invoked via the low-code interface or APIs. For example, a developer building an analytics app could plug in a NL2SQL agent to let users query a database with natural language - the agent will automatically output SQL queries and execute them, all running on the decentralized backend.

In essence, the Application Layer turns the decentralized infrastructure into a familiar “serverless” environment for AI and other applications: developers focus on high-level logic and model usage, while Ratio1 handles horizontal container placement or ChainDist job execution, concurrency, state propagation (via \textit{CSTORE}  and \textit{R1FS}), and fault tolerance. This significantly lowers the barrier to entry for AI solution deployment. Companies can leverage Ratio1 to accelerate go-to-market: instead of a large DevOps effort, they use low-code templates and Ratio1’s SDKs to stand up production-grade AI microservices in minutes.

\subsection{Components breakdown}
The Ratio1 meta-OS is architected as a multi-layered ecosystem comprising on-chain smart contracts, edge-node software agents, and supplementary decentralized services. This subsection provides an extended exposition of each core component. 

\subsubsection{Secured decentralized services} 
Decentralized orchestration in Ratio1 is realized through on-chain smart contracts and the off-chain Ratio1 Oracle network, enabling trustless management of node licensing, authentication, and workload distribution. The orchestration layer comprises:

\subsubsection*{\textit{Node Deed licensing and dAuth}}
Licensing NFTs (Node Deeds) serve as non-fungible credentials granting nodes permission to participate in the network. The decentralized authentication system (dAuth) leverages oracle-validated ownership proofs: only nodes presenting valid Node Deeds can authenticate and receive jobs. The system ensure a clear traceability between (a) node operator KYC-ed EVM compatible wallet; (b) Node Deed associated with the node operator wallet; (c) Edge Node associated with the Node Deed.

\subsubsection*{\textit{Immutable task assignment}}
Tasks are issued as transactions on the Ratio1 protocol. Every assignment, start, and completion event is recorded immutably on the blockchain, ensuring full traceability and auditability. Smart contract logic enforces service-level rules (e.g., compute time quotas, hardware verification) and automatically disburses token rewards upon proof of successful execution.

\subsubsection*{\textit{Ratio1's Deeploy API and Deeploy UI/UX}}
At the core of Ratio1 managed container orchestrations stays \textit{Ratio1 Deeploy}. Ratio1 Deeploy functions as the ecosystem’s orchestration engine, exposing a decentralized REST API for job submission and monitoring to any Deeploy API compatible dApp. Additionally, Deeploy provides a UI/UX that acts bots as a default orchestration management console as well as a template for any Cloud Service Providers willing to construct their own application. This approach dramatically simplifies and democratizes application deployment, horizontal scaling, real-time status dashboards/analytics creation and any other managed container orchestration chores for end users and administrators. Extended information on the decentralized managed container orchestration can be found in section \ref{DMCO}.

\subsubsection{Oracle network (OracleSync)} 
OracleSync is a Byzantine-fault-tolerant oracle framework that underpins system integrity by collecting node telemetry and relaying commands securely. Its principal features include:

\subsubsection*{\textit{Adapted PBFT consensus}}
Oracle nodes pool local monitoring data (e.g., CPU/GPU utilization, uptime, geographic location) from all nodes (including fellow oracles) and finalize consensus through an adapted Practical Byzantine Fault Tolerance (PBFT) protocol. This design achieves low-latency agreement without a central authority, enabling real-time trustless verification of node states. A consensus round is ran after each epoch finish resulting in a epoch finalization.

\subsubsection*{\textit{Signed computation proofs}}
Upon any important activity ranging from job completion to epoch finalization and including the constant liveness proof (or heartbeat) mechanics, Edge Nodes generate cryptographic proofs which oracles aggregate and sign. Upon consensus, these signed proofs are distributed within internal protocol blockchain, then committed on-chain to attest correct execution, preventing tampering and double-spending of compute credits. Any on-chain request reward is traceable with its cryptographic proofs and the authoring oracle nodes that participated to the consensuses.
            
\subsubsection{\textit{Ratio1} Edge Node (REN)} 

Ratio1 Edge Nodes (REN) transform heterogeneous devices-ranging from desktops and servers to mobile and virtual instances-into secure, remotely managed compute units. Key aspects of REN include:

\subsubsection*{\textit{Containerized execution environment}}
Each Ratio1 Edge Node runs as a privileged container, encapsulating dependencies and isolating workloads using both lightweight sandboxing as well as containerization itself. This design supports - besides the native application developed and deployed on Ratio1 - legacy container migration, allowing any OCI-compliant application to be deployed across the decentralized cluster, analogous to Kubernetes on the cloud-edge continuum. 

\subsubsection*{\textit{Low-Code SDK integration}}
Developers interact with Ratio1 Edge Nodes through the Ratio1 SDK, which offers high-level abstractions for job definition, data ingestion (e.g., MQTT, CSV), and result retrieval. The SDK supports multiple programming environments (Python, JavaScript, Go) and integrates with Git-backed pipelines and web frameworks to minimize operational overhead. Important to note is that executing low-code or no-code templates can be done with no costs on "paired" Edge Nodes that can be either the developer or node operator own fleet of nodes or \textit{devnet} and \textit{testnet} nodes.

\subsubsection*{\textit{Decentralized secure authentication using dAuth}}
On startup, each REN - including the oracle nodes - authenticates via dAuth against the blockchain, verifying full Node Deed ownership chain. Without a valid deed, the Edge Node will not boot, thus refusing job assignments, ensuring only licensed nodes contribute compute resources to the network.

\subsubsection*{\textit{Built-in tunneling for firewall/NAT traversal}}
By default Ratio1 ecosystem includes integration with tunneling services (e.g., ngrok) to facilitate secure connections through NATs and firewalls. This capability allows the deployment of secured, balanced, highly available, regionally balanced public or private services ranging from API endpoints to simple websites.

\subsubsection{\textit{\textit{ChainStore} (\textit{CSTORE}) - Decentralized in-memory KV-store}}
\textit{ChainStore} is a decentralized, in-memory key-value store designed to synchronize application state across multiple peer nodes with "Redis-like" usability\cite{eddelbuettel2022redis}. Rather than a single server or master, \textit{CSTORE} operates as a distributed hash table or fully-replicated data structure, so that each participating node maintains the relevant key-value dataset. This architecture eliminates any single point of failure - all nodes collectively store the state, and updates propagate peer-to-peer. In essence, \textit{ChainStore}’s design is akin to a conflict-free replicated data type (CRDT) or DAG-based store, enabling multi-master synchronization without coordination bottlenecks. Every node can accept reads/writes, and an internal consensus or gossip mechanism ensures eventual consistency among replicas. By leveraging such decentralized replication, \textit{ChainStore} remains highly available: even if some nodes go offline, the data can be served from others, and new peers can rapidly sync the in-memory state from the network.

The fully distributed nature of \textit{CSTORE} yields strong fault tolerance. There is no primary node whose failure would disrupt service - if one node goes down or loses connectivity, the remaining peers still collectively hold the complete dataset and continue serving requests. Any writes that the failed node missed can later be merged when it rejoins. This design contrasts with traditional Redis deployments that often rely on a primary/replica setup or fixed sharded clusters, which still risk downtime during leader failover or require careful configuration for redundancy. \textit{ChainStore}’s peer-to-peer replication avoids those pitfalls by design, achieving self-healing data resilience. Moreover, because each node stores (at least a portion of) the data, read operations can be served locally at each site, and write operations are disseminated in the background - an approach that can even allow near-local latency reads during network partitions, with state convergence once connectivity is restored. For example, an application using \textit{ChainStore} for session caching or feature flags would not be tied to a single cache instance - all Edge Nodes see the same logical view of the data, and transient node outages do not cause cache misses or lost updates, as the remaining nodes cover for it.

Finally, important to mention is the \textit{ChainStore} multi-tenancy and namespacing features. \textit{ChainStore} is built to securely support multiple applications or tenants on the same network. It provides isolated key namespaces (e.g., hash-based key spaces) so that different services or users can utilize the shared infrastructure without colliding or accessing each other’s data. In practice, each \textit{ChainStore} operation is scoped by a hash key or top-level namespace identifier that segregates data - for instance, a plugin or app might publish to an \verb|$hkey = “app123”| and only nodes authorized for that app will read that subset. A token-based security layer is also available on top of this. This functions similarly to a multi-tenant Redis where each application might use a separate database index or key prefix. Coupled with cryptographic access control, nodes will only replicate and serve data for authorized namespaces, ensuring that one tenant’s state does not leak to others. The Ratio1 platform enforces that nodes must explicitly “allow” each other for data sharing, establishing trust groups for \textit{ChainStore} synchronization.

By operating in a decentralized manner, \textit{ChainStore} trades some raw latency and throughput performance for \textbf{improved scalability and robustness}. A standalone in-memory cloud-native approach, being assimilated to a single-process, memory-resident database, the in-memory db can execute simple GET/SET operations in sub-millisecond time on a local server and handle millions of ops/sec on one machine. In a distributed scenario, however, a \textit{ChainStore} update involves network propagation to peers and possibly waiting for confirmation from multiple replicas, which adds overhead albeit acceptable due to efficient MQTT protocol communications.

\subsubsection{\textit{R1FS} - Ratio1 decentralized IPFS-based file system for AI artifacts}
\textit{R1FS} is Ratio1’s trustless distributed file system, built upon the IPFS protocol, that enables storage and retrieval of large model files and data artifacts across the decentralized network. It functions as a content-addressable, peer-to-peer filesystem integrated into the Ratio1 platform, providing each user with a private, encrypted storage space on the network. By leveraging a modified IPFS protocl under the hood, \textit{R1FS} can efficiently stream content fully or in chunks from multiple sources. Files added to \textit{R1FS} can be broken into content-addressed blocks and distributed among participating nodes and \textit{R1FS} relays; when a node requests a file by its content identifier (CID), the system can fetch different chunks from different peers in parallel, dramatically speeding up transfers for large files. This is especially beneficial for AI workloads - for example, a 10GB model checkpoint can be pulled to a new node by retrieving pieces from any node that already has the model, rather than a single slow origin. In essence, \textit{R1FS} uses uses a protocol akin to IPFS’s BitSwap and DHT protocols to locate the closest or fastest relay nodes holding the needed data and streams the content block-by-block.

Because \textit{R1FS} uses cryptographic hashes (CIDs) as addresses, data integrity verification is built-in - whenever a node retrieves content by CID, it can recompute the hash of the received data and confirm it matches the expected CID, guaranteeing that the content has not been tampered with or corrupted in transit

\textit{R1FS} differs from classical DFS or cloud storage in several key ways: 

\subsubsection*{\textit{(1) Decentralized control}} There is no central metadata server or single coordinator as in HDFS’s NameNode or a cloud storage provider. Instead, file continous storage is ensured by the specific subset of the oracle network and file discovery is done through a decentralized distributed hash table (DHT) announced via \textit{ChainStore}, making the system inherently peer-to-peer and removing central chokepoints. Any node can publish or retrieve content without checking in with a particular central authority but rather with the whole network of oracles, which aligns with Ratio1’s trustless model. 

\subsubsection*{\textit{(2) Content-based addressing}} Traditional filesystems locate data by hierarchy or block ID tied to specific servers, which means clients must know where data resides. \textit{R1FS} clients only need the content hash; the network of nodes cooperatively finds a copy of that content. 

\subsubsection*{\textit{(3) Built-in versioning}} If a file changes, the new version gets a new CID (since the hash changes). This means that by design, \textit{R1FS} never overwrites content in place - old versions remain addressable by their old hashes. In AI workflows, this is extremely useful for reproducibility (models or datasets can be pinned by hash) and rollback. Traditional DFS might require explicit version control or suffer from consistency issues if multiple writers overwrite a file; \textit{R1FS} avoids that by design. 

\subsubsection*{\textit{(4) Data Integrity \& Verification}} As noted, any retrieval in \textit{R1FS} double-checks the hash. In a typical distributed filesystem or network file share, clients trust the server to deliver correct data; extra checksum layers are needed for end-to-end integrity, whereas \textit{R1FS}’s integrity check is inherent to how content is addressed, thus this dramatically reduces the risk of undetected data corruption. 

\subsubsection*{\textit{(5) Peer-to-peer scaling}} \textit{R1FS} can scale with the number of nodes - particularly oracle nodes acting as \textit{R1FS} storage relays - more nodes means more copies of data and more bandwidth available for serving that data. It has no inherent single-node throughput limit because popular files will be replicated widely and served from many oracle relay endpoints.

All is based on the user-key encrypted and no single entity stores or owns the storage, thus enabling censorship-free operations and true "Your AI, your Data" approach.

\subsubsection{Encrypted Decentralized Inference and Learning (EDIL)}

Decentralized federated learning represents a cutting-edge subject and remains an active research area with significant potential for transformative applications. While our Ratio1 EDIL (Encrypted Decentralized Inference and Learning) framework demonstrates important advancements in encrypted decentralized inference and learning through novel approaches like Domain Auto-Encoder strategies and fully homomorphic encryption mechanics, we recognize this field continues to evolve rapidly. The challenges of balancing computational efficiency, privacy preservation, and scalability in trustless environments require ongoing investigation and refinement. As we advance our research and development in the area of homomorphic encryption in decentralized distributed systems and optimize the trade-offs between operational secrecy and performance, we remain committed to conducting further research and producing more advanced, incremental results that push the boundaries of what's possible in privacy-preserving decentralized machine learning.

\subsubsection*{\textit{EDIL in short}}
Our \textit{Ratio1} EDIL framework  enables homomorphic encrypted learning and serving, allowing computations on encrypted data without ever exposing the raw inputs. While traditional encryption often requires decrypting data for processing, a major vulnerability in a distributed system not to mention in a decentralized trustless ecosystem. By contrast, our proposed DFHE mechanics keep data encrypted end-to-end, significantly reducing exposure risks.

\textit{Ratio1} DFHE allows multiple nodes or participants to perform computations on the same \textit{encrypted-without-recovery} dataset without learning anything about one another’s inputs or results, making it well-suited to trustless, peer-to-peer environments. This is achieved by \textit{Ratio1} EDIL a Domain Auto-Encoder strategy: small domain-trained encoders trained via auto-encoding that compresses essential data features, functioning as a one-way encoder from the perspective of a \textit{Ratio1} training coordinator that uses its output to deliver encrypted yet usable data to each training worker. Each job type can then rely on its specialized auto-encoder, striking a balance between operational secrecy and scalability. This also mitigates the computational overhead of fully homomorphic operations and the overall overhead of training data transmission as the homomorphically encrypted data is projected into a lower dimensional embeddings space.

When deploying inference tasks, the \textit{initiator} or an external API pre-processes raw inputs using the same private domain encoder. These encrypted inputs accompany the main model to \textit{Ratio1} worker, which perform the forward pass and any specified post-processing using public open-weights models. At no point do worker nodes see the unencrypted data as Ratio1 keeps payloads encoded, thereby supporting a fully peer-to-peer inference workflow. Extra information regarding can be found in section \ref{PERF}

\subsection{On decentralized managed container orchestration - Deeploy} \label{DMCO}

Ratio1’s platform extends beyond storage to provide a managed container orchestration system, codenamed Ratio1 Deeploy, that replaces the centralized control plane of systems like Kubernetes\cite{brewer2015kubernetes} with a decentralized, trust-minimized coordination mechanism. The goal is to achieve the robust, flexible scheduling of containers across a fleet of nodes - including automated placement, scaling, and healing - without any single central scheduler or cloud provider. In Ratio1’s decentralized orchestration, blockchain smart contracts and consensus algorithms assume the role of the “control plane” that in a traditional setting would be filled by a Kubernetes\cite{brewer2015kubernetes} master or a cloud orchestrator service. This is essentially an edge-first Kubernetes\cite{brewer2015kubernetes} alternative: it can deploy containers on a network of independent nodes (edge devices, on-prem servers, cloud VMs, etc.), using the internal blockchain as well as the public (on-chain) blockchain for state synchronization and decision agreement.

\subsubsection*{\textit{Decentralized orchestration control plane}}
Rather than having one authoritative scheduler, Ratio1 employs a consensus-driven scheduling algorithm. When a user submits a deployment (for example, requesting $N$ replicas of a service container), a decentralized procedure kicks off: multiple designated orchestrator nodes (oracle nodes) will collectively determine an assignment of the $N$ containers to specific available nodes across the network. This procedure might use a combination of randomized selection, round-robin allocation, and resource availability checks to ensure fairness and load balancing. Crucially, the outcome - the mapping of containers to nodes - is recorded on the internal oracle blockchain so that it becomes tamper-evident and transparent. 

In practice, for the economics of the job distribution, Ratio1 uses escrow smart contracts to facilitate this process - namely Proof-of-AI escrow - thus when a deployment is requested, a smart contract must hold the minimal pre-payment associated with running that workload for all the future participants of the processing tasks. More information about Proof-of-AI in section \ref{POAI}.

\subsubsection*{\textit{Security and compliance considerations}}
In any production and enterprise grade system reliability, resilience, security, formal compliance are all critical \cite{babar2019security}\cite{paladi2018security} as determined by continuously improving and evolving industry standards \cite{macdonald2023cnapp}. Even more, in a decentralized orchestration, operation and interoperability with enterprise requirements (security, compliance) is paramount.  Ratio1 addresses this by using blockchain’s transparency and cryptographic identity for auditing and access control. All actions (deployments, node assignments, health signals) are logged on Ratio1 private blockchain within the oracle network as well as on-chain as hash-proofs, thus producing an immutable audit trail. This is something traditional orchestrators lack - in Kubernetes, audit logs can be tampered with by a malicious admin, whereas in Ratio1, the scheduling decisions and resource commitments are part of a public ledger. This can simplify compliance checks, as one can prove where and when a piece of data was processed or a service was running. 

Critical to note is that, for integration with identity systems, Ratio1 nodes are KYC (Know-Your-Customer) /  KYB (Know-Your-Business) verified, meaning enterprises can trust that workloads only run on nodes operated by legitimate, vetted providers (if they choose to enforce that policy via the smart contract). This is analogous to choosing specific cloud regions or approved vendors in a multi-cloud deployment. By encoding these policies in the smart contracts (for instance, only nodes with certain attributes or stake can be selected for a given job), the system can meet enterprise security rules while still being decentralized.

\subsubsection*{\textit{Ratio1 - a new standard for managed orchestration}}
Ratio1’s decentralized managed orchestration brings the benefits of cloud-native container management to a trustless environment. It achieves high availability and scalability similar to Kubernetes - automatic failover, horizontal scaling, service discovery - but does so in a heterogeneous compute continuum via peer-to-peer coordination and blockchain verification instead of through a cloud provider’s proprietary control plane. This paradigm is poised to enable \textbf{truly decentralized AI services} - for example, a globally distributed model-serving API that remains online as long as at least some subset of independent nodes agree to run it (even if others leave), with trustless consensus ensuring clients always know which nodes are responsible and that those nodes will be paid only if they perform correctly. It’s a resilient, open approach to orchestration that complements the storage solutions (\textit{ChainStore} and \textit{R1FS}) to complete our Ratio1’s decentralized AI infrastructure stack.

\subsection{System design and implementation}

Ratio1’s system design formalizes the flow from node onboarding to reward distribution\footnote{More on the formalization of PoA and PoAI in sections \ref{mining} and \ref{POAI}}.
Firstly, a participant (Ethereum account $\omega$) will get through a KYC / KYB process. After that, the user will be able to purchase a Node Deed (license token $\ell$) by sending R1 tokens to the licensing smart contract. This creates a binding $\omega \xrightarrow{p} \ell$. The user will then associates a edge device as node $\omega_n$ to $\ell$ via an on-chain transaction ($\ell \overset{\text{associate}}{\longleftrightarrow} \omega_n$), completing registration. Nodes run the \textit{Ratio1} Edge Node software and connect to the \textit{Ratio1} network. An arbitrary selected oracle - that satisfies the blockchain oracle security-cost trade-off principles \ref{eqora} where $P(\text{failure})$ represents the probability of oracle failure and $C(O)$ represents the cost of implementing oracle $O$ - running the decentralized authentication (dAuth) layer enforces that $\mathrm{owner}(\ell)=\omega$ and $(\ell \leftrightarrow \omega_n)$ hold before granting operational permissions. 

\begin{equation}\label{eqora}
 \min_{\mathcal{O}} P(\text{failure}) \text{ subject to } C(\mathcal{O}) \leq C_{\max} 
\end{equation}

Once active, nodes request and perform AI tasks dispatched by smart contracts. An authorized oracle $o$ submits cryptographic proofs of node uptime and completed work to the blockchain upon request from the \textit{Ratio1} decentralized authorized dApp. When node $\omega_n$ has valid proof, the smart contract computes the reward.

\begin{equation}
R(\omega_n) = f(\text{proof}(\omega_n))
\end{equation}

where $f(\cdot)$ is a governance-defined reward function that can be either \textit{Proof-of-Availability} or \textit{Proof-of-AI} based. The contract then executes the transfer $b(\omega) \leftarrow b(\omega) + R(\omega_n)$, crediting the owner $\omega$’s balance. We can summarize the licensing and reward flow as a tuple $(\omega,\ell,\omega_n,R)$ satisfying:

\begin{equation}
(\omega \xrightarrow{p} \ell), \ell \overset{\text{associate}}{\longleftrightarrow} \omega_n, \omega_n \overset{\text{proof($\omega_n$)}}{\longrightarrow} R(\omega_n) , R(\omega_n) \xrightarrow b(\omega)
\end{equation}

This formal chain of validation ensures that only properly licensed nodes benefit from protocol incentive\footnote{please note that PoA is based on emissions while the PoAI is entirely based on compute consumers paid fees}.

To support deployment and scalability, \textit{Ratio1} employs Infrastructure-as-Code and Software-Deployment-as-Code principles. The \textit{Ratio1} SDK lets developers define containerized AI microservices and deploy them to RENs programmatically. In practice, small-scale deployments can use the Edge Node Launcher (GUI) to automate node setup, while larger configurations use SSH-based multi-host \textit{r1setup} tool. In all cases, dAuth and oracle checks maintain security during deployment and operation.

Internally, the system leverages edge computing protocols. For example, lightweight MQTT messages trigger remote Docker containers or Python worker threads on RENs. All software components support remote configuration updates via blockchain transactions. Notably, \textit{Ratio1} extends support to chatbots and conversational interfaces: developers can deploy a Telegram bot with minimal code that interacts with on-chain AI services. This demonstrates the flexibility of the platform beyond pure ML tasks.

\section{Tokenomics and overall Ratio1 circular economic model}\label{token1}
The Ratio1 Protocol is powered by a native ERC-20 utility token (R1) used for all economic operations on the network. This utility token functions as the principal currency, facilitating payments for compute jobs, network resources, and other services throughout the ecosystem. Central to Ratio1’s design is a strictly utilitarian token model: R1 is never issued through any public sale or ICO, but instead is introduced solely through network participation (mining). These design choices ensure that token demand is driven by real usage (AI tasks, licensing, etc.) rather than speculation. Key aspects of Ratio1 utilitarian tokenomics framework are discussed below.

The complete token minting emission figures, including all ND-tier allocations, are presented in the public tokenomics worksheet\footnote{\url{https://docs.google.com/spreadsheets/d/1QmeO8-pSSZ38RnJeuWLzIyNg3VsU-MKZH0uQ34sVRxg}}.

\subsection{Pure utilitarian token}\label{token11}

The R1 token operates solely as a utility asset with no financial entitlements. It acts as the “gas” that powers protocol operations - used for processing jobs, allocating compute time, provisioning GPU resources, acquiring licenses, and similar function. Token holders receive no dividends, interest, profit shares, or other investment-style returns simply from holding R1 and, in fact, R1 is explicitly defined as a non-investment mechanism, intended only to facilitate transactions within the Ratio1 network. Unlike many web3 projects, Ratio1 has deliberately eschewed any Initial Coin Offering (ICO) or public sale and all R1 tokens (including those in the foundation treasury) are distributed via the protocol’s node-based mining process (MNDs). In short, the token value is entirely derived from its utility: it stabilizes the economy by funding AI compute and services, and it is not intended as a speculative asset.

\subsubsection{Fixed max supply}\label{token111}

The total supply of R1 is permanently capped at $N_{total} = 161,803,398$ tokens. This number is chosen as an homage to the golden ratio (1.61803398), reflecting a commitment to balanced, sustainable growth. By fixing the supply from the outset, Ratio1 ensures a transparent, finite monetary base: no additional R1 can ever be minted beyond this cap, all while any tokens entering circulation are mined through node operations. This way, economy remains strictly anti-inflationary and aligned with long-term value creation rather than arbitrary issuance. Nevertheless, due to the double-burning (20\% during ND sales and 15\% during PoAI) build-in protocol mechanics this fixed theoretical max supply is impossible to reach.

\subsubsection{Why the hard cap will never be reached:}\label{tokenBurn}

The hard cap of $N_{\text{total}} = 161{,}803{,}398$~R1 represents the \emph{theoretical} upper bound of monetary expansion, yet several built-in sinks and throttles ensure that real circulation asymptotically falls short of that figure:

\subsubsection*{\textit{(1) Burn on licence issuance}}
Whenever a public Node Deed (ND) is purchased, the licence contract burns a fixed fraction of the payment.  
Those units are destroyed permanently, reducing the tranche that would otherwise vest to the node.

\subsubsection*{\textit{(2) Burn on Proof-of-AI execution fees}}
For every PoAI job, \textbf{15\,\%} of the escrowed fee is sent to the null address.  
Because PoAI volume scales with genuine network usage, this burn offsets new emissions as adoption grows.

\subsubsection*{\textit{(3) Tiered ND pricing (anti-inflation schedule}}
ND licences are issued in successive tiers, each priced higher than the last.  
Unsold tiers correspond to un-minted rewards; their associated emission can never occur.

\subsubsection*{\textit{(4) Unallocated Master Node Deeds}}
A portion of MND slots is reserved for future strategic partners and stewards.  
Until such deeds are minted on-chain, their reward streams remain un-minted.

\subsubsection*{\textit{(5) Inactive or orphaned licences}}
Even after minting, an ND or MND begins to earn only when it is \emph{coupled} to a live, KYC-verified server via the dAuth interface.  
Licences that are never staked to hardware simply mint nothing, permanently lowering potential supply.

\subsubsection*{Simulations}

Some realistic token simulations emission figures based on possible adoption, minting rewards for both PoA and PoAI, burn, including ND-tier allocations and allocated MND, are presented in the public tokenomics worksheet - second tab\footnote{\url{https://docs.google.com/spreadsheets/d/1QmeO8-pSSZ38RnJeuWLzIyNg3VsU-MKZH0uQ34sVRxg}}.

\subsubsection{Mining via licensed Edge Nodes PoA}\label{token112}

In the Ratio1 ecosystem, new tokens can only be generated by operating licensed Edge Nodes. Each participant must acquire a Node Deed (ND) license and run a functional node; then the network’s hybrid consensus mechanism distributes mining rewards based on node performance. Specifically, token rewards are earned via Proof-of-Availability mining (ensuring the node stays online and available) as well as combined with Proof-of-AI  (verifying useful AI computation) fees. In effect, every licensed node contributes computing power and is compensated in R1 - there are no pre-mines or ICO distributions beyond this scheme. This approach guarantees that R1 issuance is fully transparent and reflects actual network contribution.

\subsubsection{Licenses}\label{token113}

Ratio1 employs three classes of licenses for its Edge Nodes. The smart-contract bound licenses come in three classes of node deeds: Genesis Node Dee (GND), Master Node Deed (MND), and standard Node Deeds (ND). Each class grants different privileges, token allocations, and emission schedules, as detailed in the subsections below.

\subsubsection*{Genesis Node Deed (GND)} 
The Genesis Node Deed is a single, foundation-held license that kickstarts the protocol’s compute network. Activating the GND brings online the first “Edge Node zero” and begins minting tokens for the system. It serves as the cornerstone of the ecosystem’s launch: the GND controls about 28.9\% of the total R1 supply, which will be emitted over roughly 12 months beginning with the protocol’s genesis on 23 May 2025. Tokens generated by the GND are allocated to publicly disclosed foundation pools supporting development and community initiatives. More information about these pools in section \ref{WALLETS}

\subsubsection*{Master Node Deed (MND)} 
Master Node Deeds are special licenses reserved for the founding team and select seed investors. These nodes typically operate oracle services and other critical infrastructure, and their rewards vest over a long schedule (approximately 30 months starting shortly after launch). Importantly, MNDs themselves do not earn Proof-of-AI rewards: by protocol design, MND holders contribute computation free to support the network rather than generate AI-work tokens - If founders want to perform paid AI jobs, they must also acquire a standard Node Deed. Overall, MND licenses account for roughly 26.1\% of R1, reflecting the founders’ vested stake in the platform.

All MND emissions follow an \textbf{sigmoid emission schedule curve}: after a seven-month
cliff a small fraction begins to vest, the release rate then accelerates sharply until roughly
month~22, and finally tapers so that distribution completes near month~30.
This \emph{S-shaped} schedule discourages early sell-offs, keeps founders and
seed partners aligned over the long term, and suppresses the late-stage
inflation that a simple linear vesting model can introduce.

\subsubsection*{Node Deed (ND)}\label{token12}
Standard Node Deeds are public licences that anyone can buy with R1 tokens, giving them the right to operate a Ratio1 Edge Node. Each ND grants the right to mine up to 1,575.19 R1 tokens, which are released gradually over a 36-month period once the node is active (and stays active). Node Deed operators earn rewards both for keeping their nodes online (Proof-of-Availability mining) and for executing paid AI compute jobs (Proof-of-AI fees). In aggregate, NDs PoA mining represent about 45\% of the total R1 supply, making them the primary mechanism for public participation: any node owner can contribute hardware or cloud resources and be compensated in R1 for supporting the network’s AI training, inference, and other services.

Upon every ND purchase the protocol auto-routes the payment:\; \textbf{50\,\%} is added to the liquidity LP wallet, \textbf{20\,\%} is irreversibly burned, and the remaining \textbf{30\,\%} is transferred to the operational-expenses wallet.

\textit{Secondary-market transparency.}  
The Ratio1 Explorer \footnote{\url{https://explorer.ratio1.ai}} displays, for every ND,
(i) the number of PoA tokens already minted out of 1 575 and  
(ii) the residual emission eligible to the current holder.
Prospective buyers can therefore price an ND precisely.  
PoAI income remains uncapped and is determined by future protocol
utilisation, as detailed in Section~\ref{POAI}.

\subsection{Proof-of-Availability rewards}

Before diving into Proof-of-Availability, note that PoA rewards are not passive income; they are active mining yields that cease if the operator stops maintaining the node. This is a major potential misunderstanding and it is entirely invalidated by the fact that without the Node Operator care and basic operation services the Ratio1 Edge Nodes might discontinue the services thus leading to decrease and even nullifying the rewards.

\subsubsection{Token mining approach}
The Ratio1.ai Proof-of-Availability rewards constitute a novel utility token mining mechanism that fundamentally diverges from traditional computational proof systems by establishing consensus-based availability assessment as the primary determinant of token distribution.  This approach represents a significant advancement in distributed ledger technology, where the mining process is intrinsically linked to the verifiable uptime and operational capacity of individual network participants rather than computational resource expenditure. The underlying architecture operates on discrete temporal intervals, specifically 24-hour epochs, during which each participating node maintains continuous communication with the network infrastructure through automated liveness-proof emissions. These liveness proofs serve a dual purpose: first, they provide cryptographic evidence of node operational status, and second, they convey real-time job loading information that enables dynamic resource allocation optimization across the distributed network.

\subsubsection*{\textit{Illustrative and formal analysis of the minting timeline}}\label{mintTimeline}
Assume a standard Node Deed (ie. not a MND nor the ecosystem GND) is activated on \textbf{1 Jan 2026}.  
As stated in Section~\ref{token113}, an ND can mint at most 1\,575 R1 over
$36\times30\!=\!1\,080$ daily epochs, but \emph{only} while the associated
Ratio1 Edge Node maintains availability above the 98 \% threshold further presented and explained in section \ref{mining}.  
Because the emission clock pauses during downtime, the wall-clock end date
depends on the operator’s behaviour as in next example:
{2026} - node uptime $=100\%$; first 360 epochs complete, so 525 R1 (one third of the cap) are minted.
{2027} - node offline for the entire year; zero epochs accrue, so no PoA rewards minted.
{2028--2029} - uptime restored to $100\%$; the remaining 720 epochs elapse over two calendar years, so the final token mints on \textbf{1 Jan 2030}.

Hence “three years of minting” means \emph{three years of verified
availability}, not three consecutive calendar years.

Formally, in order to better understand the Node Deed mining schedule, we can model a ND license as entitling its \textit{R1OP} to a fixed total of R1 tokens, minted linearly over a minimal number of daily epochs (a 36‑month period). Each epoch $e$ is a 24‑hour interval during which a node $k$ uptime is measured. Let \(N=1575.19\)\,R1 be the fixed minting cap of a Node Deed and
\(T=1080\) the number of \emph{epoch-credits} to be accumulated.
Each epoch \(e\in\mathbb N\) is a 24\,h window.
Denote by \(a_{k_e}\in[0,1]\) the fraction of that window during which the node $k$ is verifiably online, measured via 10–15\,s heartbeat proofs. For service-level analytics we also record the Boolean flag
\(S_e=\mathbf 1(a_{k_e}\ge0.98)\), but \(S_e\) does not affect payouts.

\paragraph{Per-epoch mint.}
\begin{equation}
  r_{k_e} \;=\; \frac{N}{T}\,a_{k_e} .
  \label{eq:epoch-reward}
\end{equation}

\paragraph{Cumulative state after \(m\) calendar days.}
\begin{equation}
  R(m)=\sum_{e=1}^{m} r_{k_e}
       \;=\; \frac{N}{T}\sum_{e=1}^{m} a_{k_e},
  \quad
  C(m)=\sum_{e=1}^{m} a_{k_e} .
  \label{eq:cumulative}
\end{equation}

\paragraph{Completion time.}

\begin{equation}\label{miningfinish}
  C(m)\;\ge\;T,
  \qquad
  R(m)=N.
\end{equation}

\paragraph{Expected mining}
\begin{equation}\label{expectedmining}
E[t_\star]=\frac{T}{\bar a}, \text{ given average availability
} \bar a
\end{equation}

Minting - ie. Proof-of-AI mining - ceases on the first day when the equation \ref{miningfinish} is satisfied.
Given an average availability $\bar a$, the expected horizon is
denoted by equation \ref{expectedmining}.
Thus perfect uptime ($\bar a=1$) yields exactly 1080 days, whereas
$\bar a=0.90$ extends the schedule to roughly 1200 days.

This proportional-accrual design both caps the total supply and rewards
every verified second of service, while the 98 \% flag \(S_k\) remains
available for external reputation or slashing mechanisms.

More information about Proof-of-Availability mining rewards formalism wrt. oracle network consensus can be found below in section \ref{mining}

\subsubsection{Oracle nodes revisited}

The supervisory framework employs a specialized class of network participants designated as Ratio1 oracle nodes, which function as elevated-trust entities with enhanced computational and monitoring responsibilities. These oracle nodes implement a comprehensive data capture, verification, and storage protocol that ensures the integrity and completeness of availability metrics across all network participants. Important to note here is that the oracle architecture represents a critical departure from traditional peer-to-peer consensus models by introducing hierarchical oversight mechanisms that enhance both security and operational efficiency.

The oracle network's supervisory role extends beyond passive monitoring to encompass active verification of node performance metrics, job completion rates, and network contribution assessments. Thus, this multi-dimensional evaluation framework ensures that availability measurements reflect not merely uptime but genuine network utility and resource contribution.

Each oracle node operates under either an MND or an ND licence, and no two nodes are controlled by the same private or public entity. Because Ratio1 maintains a clear, on-chain relationship and identification between every Edge Node, its MND/ND licence, and the KYC/KYB-verified owner of that licence (EOA), it is fully transparent and easily verifiable that no single entity owns or manages more than one oracle.

\subsubsection{Adapted Practical Byzantine Fault Tolerance Implementation}

Following epoch completion, the oracle network initiates a sophisticated consensus process based on an adapted Practical Byzantine Fault Tolerance (apBFT) algorithm specifically optimized for availability assessment in distributed edge computing environments. The apBFT implementation incorporates self-exclusion and reputation-based supervisor (oracle) node selection mechanisms and dynamic timeout adjustments to enhance consensus reliability while maintaining resistance to Byzantine attacks. 

The consensus process employs a multi-phase protocol where oracles aggregate individual node availability assessments through cryptographically secured voting mechanisms using the Ratio1 internal communication and decentralized storage services. The final goal of this approach is to ensures that even in the presence of malicious or faulty oracle nodes, the network can achieve reliable consensus on node availability metrics, provided that fewer than one-third of the oracle nodes exhibit Byzantine behavior \footnote{Further information regarding the internals of our aPBFT protocol and its inter-operation with other ecosystem components will be published in our series of research papers}.

\subsubsection{Token mining formalism}\label{mining}
The Proof-of-Availability reward function $R_{PoA}(k,e)$ provides the reward quantity or R1 tokens for a Ratio1 Edge Node $k$ at epoch $e$ based on the oracle network computed availability percentage $A_{apBFT}(k, e)$. $A_{apBFT}(k, e)$ in turn, resulted from the apBFT consensus of the oracle network w.r.t. the availability of Ratio1 Edge Node $k$ at epoch $e$. This formulation ensures that token distribution remains directly proportional to verified network contribution while maintaining cryptographic verifiability of all reward calculations, while the consensus-derived metric provides a robust foundation for token allocation that resists manipulation and ensures fair distribution based on genuine network participation.

The consensus resulted availability $A_{apBFT}(k, e)$ is based on $H_{O_i}$, ie the byte-normalized value between 0 and 255 of liveness-proofs that each individual oracle $O_i$ received from the said node $k$ during epoch $e$. This $[0..255]$ normalization process ensures consistent measurement scales across diverse hardware configurations and network conditions while maintaining cryptographic integrity of the underlying proof data.

Each oracle $O_i$, out of the total population of $O$ oracles, computes $H_{O_i}(k,e)$ - its individually computed Edge Node $k$ availability at epoch $e$ - before participating in the consensus and will not propose its state unless the self-availability assessment results in a value above a configured threshold. The individual oracle computation takes into consideration the set $H_{k_e}$ of of liveness proofs received during epoch $e$ from node $k$ versus the ideal quantity and quality of this set $Mh_{k_e}$. As mentioned before, this percentage is normalized between 0 and 255, thus we denote $\alpha = 255$.
As such, the consensus protocol incorporates a sophisticated threshold mechanism whereby oracle nodes are required to achieve a minimum self-availability assessment before participating in the consensus process for other network nodes - specifically, each oracle must demonstrate availability levels exceeding a configured threshold, set at 98\%, before its availability assessments of other nodes are considered valid for consensus participation. This threshold-based participation model serves multiple critical functions: it ensures that only highly available and reliable oracle nodes contribute to consensus decisions, it prevents degraded oracles from introducing bias into availability assessments, and it maintains the overall integrity of the consensus process even during periods of network stress or targeted attacks.

Below is the summarized mathematical formalization of the rewards calculation considering the already notations presented in the equations from section \ref{mintTimeline}.

\begin{equation}
H_{O_i}(k, e) = \lfloor \frac{\lvert H_{k_e}\rvert}{Mh_{k_e}} \times \alpha \rfloor 
\end{equation}
\\

\begin{equation}
V_{H(k,e)} \;:=\; \bigl(H_{O_i}(k, e)\bigr)_{i=1}^{O}, \text{where } H_{O_i}(k,e) \in \{0,1,\dots,255\}
\end{equation}
\\

\begin{equation}
A_{apBFT}(k, e) = \operatorname{med_{apBFT}}^{*}({V_{H(k,e)}})
\end{equation}
\\

\begin{equation}
R_{PoA}(k, e) = \frac{N}{T} \times \frac{A_{apBFT}(k, e)}{\alpha}, \text{where }  A_{apBFT}(k, e) \in \{0,1,\dots,255\}, 
\end{equation}

\subsubsection{Proof-of-AI Rewards}\label{POAI}

A key innovation in Ratio1’s orchestration is aligning economic incentives with reliable service delivery. Node operators - basically decentralized cloud providers - are rewarded via the R1 token economy \textbf{fees} for contributing compute, but only if they perform as agreed. 

When a node is selected to run a container or a native application job, an escrow contract locks up a reward (paid by the developer through the CSP) for that node, conditional on it maintaining the service for a specified epoch duration. As an example, Ratio1’s mainnet uses 30 epochs as the period a node must stay online with the service. Any downtime below this threshold generates penalties, node score penalty and job transfer to another node.

Thus, if the node successfully meets the uptime and performance criteria throughout that period, the contract releases the payment to the node at the end, with a portion (ie 15\%) being burned (ie. send to null address) as ecosystem gas fee. Important to note is that \textbf{0\,\%} is retained by the genesis company (zero commission) and \textbf{85\,\%} paid directly to the R1OPs that had their processing nodes involved in the job.

Concretly, when an \textit{Ratio1} Edge Node executes a task, it contributes resources such as CPU/GPU cycles, memory, and storage. The protocol quantifies the Edge Node $k$ resource allocation $C_{e}(j)$ (e.g., vcpu and memory allocation) at epoch $e$ for a job $j$ and issues a reward $R_{e}(k)$ in the R1 utility-token proportional to the availability of the node $k$ and the complexity of the give. Formalized the PoAI reward can be described as:
\begin{equation}
    R_{e}(k) = \sum_{J}f_{PoAI}(C_{e}(j), A_{k}(e))
\end{equation}
Thus, nodes that supply more computation receive proportionally more tokens. These rewards are effectively "mined" by useful work, establishing an authentic AI-mining system. An escrow smart contract  holds payment until oracles verify task completion, then releases R in R1 tokens to the node on demand. 

\subsection{Ecosystem treasury and pools} \label{WALLETS}

The Ratio1 treasury is divided into dedicated pools aligned with distinct strategic objectives. The genesis company, as the genesis governing entity, holds separate multisig wallets for each pool (LP, research/R\&D, marketing/community, hackathons/grants, and social responsibility) that are publicly listed on-chain. All token allocations are predefined and transparent: there was no premine, and all emissions occur on-chain via node-deed mining (PoA), meaning that every token in these pools is earned by network contribution of the Ratio1 seed nodes. Each pool’s disbursements follow strict governance: multisig approval, milestone-based releases, and public reporting are required.  In fact, the Ratio1 Foundation has pledged a “Zero Token Consumption” promise - it will not distribute or sell any tokens from the Marketing, Grants, or CSR pools (together ~59.4\% of Genesis mining rewards) until the protocol reaches advanced maturity - at least 12 months as well as multiple internal KPI related to adoption and job consumptions\footnote{More information about the internal KPI will be further disclosed in the Ratio1 Blog}. All these wallets and the movements are compiled in the Ratio1 explorer dApp\footnote{https://explorer.ratio1.ai} and fully verifiable via the on-chain explorers.

\subsubsection*{LP Wallet}
Exclusive Liquidity Provision for Decentralized and Centralized Exchanges, the LP Wallet receive R1 mining rewards and maintains a singular focus on providing healthy token liquidity across trading platforms without engaging in speculative activities. This approach ensures stable market conditions while avoiding manipulation tactics commonly seen in cryptocurrency projects. This wallet receives 26.7\% of the Genesis Node Deed mining rewards - about 7.7\% of the total protocol.

\subsubsection*{Marketing Wallet}
Together with the Grants and the CSR Wallet, the Marketing wallet will remain dormant until Ratio1 achieves sufficient business traction and demonstrates measurable utility KPIs in terms of both decentralized compute and storage providers but more importantly compute consumers. This share of 7.5\% of the overall Genesis Node Deed mining rewards (about 2.2\% of the protocol) will be used for marketing investments.

\subsubsection*{Grants}
Probably the most important wallet in terms of community and development impact, summing 34.6\% of the whole Genesis Node Deed mining rewards allocation - the biggest share - and about 10\% of the whole protocol allocation - about 16M R1 tokens. This critical wallet will also be "unlocked" after important traction indicators are reached and will enable the funding of projects build on top of the Ratio1 protocol. Ratio1 Grants mission is to sponsor developer hackathon events (often with technical support and prize funding) to crowdsource innovative applications on the platform. Promising teams can then receive continued grant funding to integrate their solutions into the Ratio1 marketplace. The genesis company and the community of stakeholders oversees grant disbursements via milestone-based agreements: recipients must meet development milestones (code deliverables, open-source releases, etc.) to draw funds. Throughout, transparency is enforced: grant calls, winner announcements, and post-mortem reports are published, ensuring the community sees exactly how these resources are used. This model aligns with Ratio1’s ethos of community-led innovation, as it channels treasury value directly to contributors who enhance the ecosystem

\subsubsection*{Social Responsibility Pool}
Dedicated to ethical engagement and impact, this pool funds initiatives that “give back” in line with Ratio1’s values. Typical uses include educational outreach (e.g.,\ free AI training workshops), sustainability projects (e.g.,\ energy-efficient computing programs), and other social-good efforts. About 17.3\% of Genesis mining rewards (~5\% of total supply) were reserved in the CSR wallet. These funds are locked until late-stage maturity; as noted already, the CSR wallet will be unlocked last, specifically to fund social projects once the protocol has proven its success. Any CSR spending must adhere to predefined ethical guidelines (ie. environmental or community criteria) and is subject to the same multisig and reporting rules: detailed impact plans and outcomes will be documented publicly. By allocating a dedicated CSR fund with these oversight measures, Ratio1 embeds its commitment to ethical AI and societal benefit into the very fabric of its treasury.

\section{Performance and security consideration}\label{PERF}
Ratio1 is designed for high performance and scalability through parallelism. In an ideal scenario, if a task can be divided among $n$ independent RENs, the total execution time decreases roughly as $1/n$. Formally, if $T(n)$ is the latency with 
n nodes and $T(1)$ with one node, then the speedup $S(N)=T(1)/T(n) \approx n$. Actual speedup will be somewhat lower due to communication and coordination overhead. \

In the \textit{Ratio1} Decentralized Federate Homomorphic Encryption pipeline, we factor the model into encoder and worker components. As shown in our research, the end-to-end complexity of the model $M$ satisfies:

\begin{equation}
    O(M) = O(E) + O(W)
\end{equation}
where $O(E)$ is the complexity of the private encoder and $O(W)$ is the the complexity of the worker’s network. For efficiency we ensure $O(E) << O(W)$, meaning the overhead of encryption (encoding) is minimal compared to the heavy training or inference work. In practice, our experiments with encrypted inference on latent representations indicate only a small slowdown (e.g., under 10\%) relative to unencrypted pipelines. Empirical validation on test networks confirms robust scaling. In a synthetic training workload distributed over 10 RENs, we observed near-linear throughput increase, consistent with our theoretical model. For latency-sensitive inference, the decentralized system achieves comparable end-to-end delays to a small-cloud setup when a sufficient number of RENs is used. Overall, \textit{Ratio1}’s architecture allows it to scale horizontally with the number of active nodes, while overheads from on-chain communication and encryption remain bounded by design.

Security is built into \textit{Ratio1} at every layer. Data confidentiality in decentralized MLOps is addressed via the Decentralized Federated Homomorphic Encryption (DFHE) mechanism. Under DFHE, raw inputs $x$ are first mapped by a private encoder $E:X\rightarrow Z$ into encrypted latent space. Workers then train or infer using only 
$z=E(x)$, never seeing the original $x$. Mathematically, the end-to-end model is 
\begin{equation}
    M=W \circ E    
\end{equation}

where $W$ (the worker model) operates on the latent space. Since the decoder is withheld, z remains non-invertible to workers. This preserves privacy even if workers are malicious. The formal protocol ensures all training and inference computations occur on encoded data; only the trusted master (node owner) holds the encoder. As a result, the system effectively has full end-to-end encryption for ML tasks. Future work includes integrating zero-knowledge proof schemes, to allow a new appraoch for workers to prove correct computation.

On the blockchain side, all smart contracts (licensing, escrow, etc.) are designed for formal analysis. By using immutable on-chain code and well-known primitives, we minimize the attack surface. The \textit{Ratio1} implementation can be subjected to standard smart contract audits and formal verification tools (e.g., model checking or Coq proofs) to ensure correctness. \textit{Ratio1} Edge Nodes and all the services communicate using encrypted and authenticated through digital signatures messages. Each REN generates a keypair and the node operator confirms its public key on-chain via the \textit{Ratio1} dApp then in turn the dAuth system ensures the automated authentication. All inter-node communications (e.g., via MQTT) include signed metadata so that any tampering is detected. The underlying blockchain provides immutability for all job logs and payment transactions, preventing fraudulent activity. In summary, the protocol uses a combination of encryption, consensus, and cryptographic authentication to maintain a trust-minimized environment for decentralized AI.

\section{Team and Technical Expertise}

The \textit{Ratio1} project is spearheaded by a multidisciplinary team of distinguished researchers and engineers affiliated with \textit{Ratio1}.ai and its partner institutions. The core authors of this work, including A. I. Damian, C. Bleotiu, M. Grigoras, P. Butusina, A. De Franceschi, and V. Toderian, collectively embody a wealth of expertise spanning critical domains in modern heterogeneous distributed computing and artificial intelligence. 

Each team member brings a robust background in either blockchain technology or data science related fields - advanced data analysis, statistical modeling, and predictive analytics - complemented by extensive experience in designing and implementing scalable, fault-tolerant distributed computing architectures. Their proficiency extends to the development and deployment of sophisticated machine learning models and pipelines, underpinned by rigorous DevOps and MLOps practices that ensure seamless automation of software development, deployment, and monitoring processes, including continuous integration and delivery.

The team demonstrates also particular strength in smart contract engineering, with proven capabilities in writing, testing, and deploying smart contracts on various blockchain platforms. Their expertise in cloud infrastructure design and management enables the development of scalable applications that can handle enterprise-level workloads while maintaining optimal performance and reliability.

\textbf{Andrei Ionut Damian, Ph.D. (CEO).} Dr.\ Damian is a veteran AI researcher and serial entrepreneur with over 25 years of experience in artificial intelligence.  He earned his Ph.D.\ in computer science/AI and serves as a lecturer in data science at the Polytechnic University of Bucharest.  His career spans both academia and industry: he has co-founded AI-driven technology companies and published research on neural architectures and machine learning operations.  Under his leadership, Ratio1 integrates advanced distributed AI infrastructure with blockchain-based coordination, reflecting his technical expertise in large-scale, secure AI systems.

\textbf{Cristian Bleotiu (Data Scientist).} Bleotiu is a computer science graduate specializing in natural language processing.  He has co-authored research on distributed AI platforms, notably contributing to the previous team published research.  Within Ratio1, he focuses on the development of AI models and NLP components - including Foundation models \textit{J33VES} framework - leveraging his background in machine learning and competitive programming to advance the platform’s intelligent services.

\textbf{Marius Grigoras (Senior Technical Leader).} Grigoras is an experienced software engineer and systems architect, with more than 15 years in computer science and a specialization in security, embedded systems, and distributed networks.  He is co-founder and CEO of BHero Network, a blockchain, security and AI infrastructure startup.  At Ratio1 he provides technical leadership for infrastructure and blockchain integration, guiding the design of the decentralized cloud-on-edge architecture in line with his expertise in scalable, secure and secure distributed systems.

\textbf{Petrica Butusina (Senior Product Owner).} Butusina is a seasoned product strategist and designer with extensive experience in blockchain technology and digital products.  He co-founded and leads product development for BHero and FLIPiX (blockchain-based projects) and has worked on high-profile Web3 and NFT initiatives.  In the Ratio1 team, he combines his skills in product development, blockchain economics, and community engagement to shape the platform’s ecosystem and user-facing offerings.

\textbf{Alessandro De Franceschi (Senior Software Engineer).} De Franceschi is a full-stack developer specializing in Web3 and blockchain technologies.  He has deep expertise in smart contract development as well as frontend and backend engineering, with a focus on security and scalability.  Active in the blockchain space since 2020, he applies this technical skill set to build Ratio1’s decentralized software components and secure system integrations.

\textbf{Vitalii Toderian (Machine Learning Engineer).} Toderian is a machine learning specialist completing an undergraduate degree in Computer Science at UPB and pursuing graduate studies in AI/Data Science.  He contributes to the development of the Ratio1 AI stack, applying machine learning techniques to on-device and distributed AI services.  Under the mentorship of senior researchers, he works on implementing ML pipelines and intelligent agents for the platform, leveraging his academic training and research interests in AI.

\section{Conclusion and future tech roadmap outlook}

Ratio1 implements a new disruptive approach that unifies decentralized computing and machine learning orchestration.  In this paradigm, everyday devices (from laptops and tablets to servers) become Edge Nodes that jointly host AI workloads, effectively creating a global "compute ride-sharing" network.  This meta-OS replaces traditional centralized ML pipelines with a peer-to-peer architecture: tasks are scheduled across independent nodes under smart-contract coordination, and compute resources are shared dynamically in exchange for tokenized micropayments.

\subsection{AI meta-OS}  

Crucially, Ratio1’s design integrates several key innovations. Oracle-based coordination (via the \textit{R1 OracleSync} module) provides trustless task validation, decentralized storage layers (\textit{R1FS and \textit{CSTORE}}) prevent any single point of failure, and a decentralized authentication layer (dAuth) replaces centralized credentials.  A \textit{low-code workflow} framework lets developers compose AI applications from high-level templates with minimal coding effort, while an on-chain licensing model (ERC-721 "Node Deeds") formalizes node participation.  Together with advanced privacy measures-such as homomorphic encryption (the \textit{"EDIL”} architecture) that allows inference on encrypted data, all this stack forms a \textbf{holistic and extensible foundation for next-generation AI services} and more.  In short, Ratio1’s meta-OS moves beyond isolated innovations to offer a \textbf{comprehensive decentralized MLOps platform} where compute is shared, verified by oracles, and paid for via smart contracts, paving the way for more scalable and robust AI pipelines.

Ratio1’s modular architecture can be seen as a toolkit for building distributed AI applications.  Core components like R1FS, \textit{CSTORE} enable seamless data sharing and state synchronization across nodes.  The system’s token economy aligns incentives: participants \textit{buy node licenses} on-chain and earn R1 tokens by keeping their nodes online and processing tasks.  Oracles sign off on node availability and task completion, enabling \textit{Proof-of-Availability} consensus that ensure fairness and reliability in reward distribution.  Meanwhile, a smart-contract-based orchestration layer handles licensing, escrows, and micro-payments automatically.  In sum, the Ratio1 meta-OS not only unifies decentralized compute and storage, but also embeds a full-stack coordination layer (oracles, contracts, encryption) and a user-friendly interface, addressing many gaps that exist in prior distributed MLOps efforts.

\subsection{AI democratization}  A core goal of Ratio1 is to \textbf{democratize AI} by lowering infrastructure and expertise barriers.  The platform explicitly targets broad accessibility and the vision is that "if you have a device with spare cycles, you can contribute it; if you have an idea for an AI app, you can \textbf{develop as well as deploy} it without owning hardware".  In practice, this means advanced ML can run on consumer-grade devices rather than proprietary data centers.  By leveraging blockchain-based micropayments and token incentives, Ratio1 makes compute affordable - only lightweight client nodes and applications are needed, and participants pay each other in small R1 transactions rather than sinking capital into cloud servers.  This shifts the economics of AI: the "AI power to the people" paradigm allows any user (novice or expert) to tap into the collective pool of models and data without owning a supercomputer.

In doing so, Ratio1 opens new revenue models and participation pathways.  AI developers, researchers, and even hobbyists can run or deploy models with minimal DevOps overhead, thanks to the low-code tools and the public node network.  Conversely, \textbf{node operators} earn R1 tokens by contributing idle resources to real-world ML tasks or cloud-on-edge services.  This mutual marketplace - reminiscent of Uber or Airbnb but for compute - means that an AI service consumer and a resource provider become equally valued network participants.  Importantly, this democratization extends beyond technical access: by lowering costs and complexity, Ratio1 aims to include users from diverse backgrounds.  As the developers emphasize, universal access to AI  - ubiquitously available across computers and mobile devices globally - is the long-term vision.  By making machine learning infrastructure lightweight, tokenized, and decentralized, the ecosystem empowers communities and small organizations to innovate with AI where only large corporations once could.

\subsection{Further technical improvements}  

\subsubsection{Training and serving}
We are actively advancing the Ratio1 protocol in multiple technical dimensions.  Performance engineering remains a priority: for instance, optimizing the \textbf{EDIL}  layer is critical so that encrypted inference and training become practical on edge hardware.   On the deployment side, we plan improving the heterogeneity of the nodes (GPUs, ARM-based devices, specialized AI chips, etc.) to validate performance under real-world load on \textbf{any} device.  This includes porting the Ratio1 runtime to non-x86 platforms (ARM, RISC-V, etc.) and ensuring compatibility with a broad range of hardware profiles and operating systems.  We will also continually integrate new machine learning models as they become available: the \textit{J33VES} framework is designed to support any open-weight LLM or other open models, so future breakthroughs in on-device AI (e.g.,\ LLMs, vision models, or lightweight robotics perception models) can be plugged in seamlessly.

\subsubsection{Security, safety and resilience}

Other upcoming priorities are \textbf{security, safety and resilience}.  In line with emerging regulatory standards, we align the managed orchestration and application layers of Ratio1 with frameworks like the EU’s Digital Operational Resilience Act (DORA) \cite{eu_dora_2022}.  This means building in robust ICT risk management, incident detection and reporting, and continuity practices that financial institutions and critical services will demand.  For example, we will harden the Edge Node environment (e.g.,\ through secure-boot, code-signing, and integrity attestation) because, as noted in our trust-analysis, consumer-grade nodes can be tampered with if unprotected.  We will also continue fortifying the data infrastructure: the decentralized file-system (R1FS) and the ChainStore/\textit{CSTORE} services have built-in redundancy and content-addressing for integrity, but we aim to improve their reliability further (for instance via additional replication, integrity checks, and faster recovery mechanisms).  In short, we treat resilience and security as first-class citizens: the system must not only operate in the open, but do so reliably and predictably under adverse conditions, much like a regulated cloud provider.

\subsection{Community network}
Finally, our internal road-map includes governance and sustainability milestones. We envision evolving toward a more \textbf{community-driven governance model} where active stakeholders (e.g.,\ licensed node operators, service providers, developers) can have formal influence.  This could take the form of on-chain governance or collaborative decision bodies that “bring stakeholders into the loop” as the network grows.  Such an approach would align with our broader goal of a self-sustaining ecosystem.  Notably, by design Ratio1 already promotes environmental sustainability: pooling idle compute resources reduces wasteful cloud usage, which research suggests can significantly cut AI’s carbon footprint.  Going forward, we will continue to optimize for efficiency (for example by dynamic power management and workload scheduling) to further this aim.

Ultimately, Ratio1’s long-term objective is a \textbf{sustainable, inclusive AI ecosystem}. By coupling cutting-edge research with real-world engineering, we are laying the groundwork for an AI paradigm that is both technologically advanced and broadly accessible.  As our cited work describes, this convergence of decentralized computing and AI "could inspire future innovations" and pave the way for AI solutions that operate at the edge, respect user privacy, and reward contributors equitably.  We are committed to iterating on this vision, refining each component of the stack, and working with the community to realize a resilient "AI for everyone" future.

\section{Disclaimer}\label{disclaimer}

The information in this White Paper is conceptual and subject to ongoing
legal, regulatory, tax, technical, and compliance reviews.  We reserve the
right to modify any part of it as market conditions and professional advice
may require.  This document is not a prospectus, nor does it constitute an
offer of financial instruments, investments, or other regulated products or
services.

\vspace{0.5em}
\noindent\textbf{Issuer.}
The issuer of the Node Deeds, Master Node Deeds, and Genesis Deed that mine
R1 Tokens (the ``Tokens'') is \textsc{Naeural SRL}, a Romanian company with
registered office at \emph{61 Eufrosina Popescu St.\, B 42, District 3,
Bucharest}, registered under J40/12133/18.06.2024, fiscal code 50252500
(hereinafter the ``Company'').

\vspace{0.5em}
\noindent\textbf{No public sale.}
Tokens are \emph{not} issued via an ICO or public sale; they enter
circulation solely through mining, ensuring demand is driven by real usage.

\vspace{0.5em}
\noindent\textbf{Not securities.}
Tokens, Node Deeds, Master Node Deeds, and the Genesis Deed are utility
items and do not constitute securities, financial instruments, or
investments.  They convey no ownership, voting rights, or profit share in
the Company.

\vspace{0.5em}
\noindent\textbf{Utility licences.}
Node Deeds function as utility licences: any Tokens mined depend on the
holder’s active operation of a node.  Rewards are earned through the
holder’s own computational effort within a decentralised network and are
not passive returns managed by the Company.

\vspace{0.5em}
\noindent\textbf{Regulatory position.}
Because Node Deeds are utility licences, their sale and operation do not
require the Company to hold securities licences in the EU or USA.  Buyers
are responsible for compliance with any local laws that apply to their
purchase or use of Node Deeds.

\vspace{0.5em}
\noindent\textbf{No advice.}
Nothing in this White Paper constitutes financial, investment, legal, or
tax advice, nor a solicitation to buy or sell Tokens or Node Deeds.  Readers
should obtain their own professional advice.

\vspace{0.5em}
\noindent\textbf{Risk acknowledgement.}
Holders acknowledge risks relating to blockchain technology, loss of
private keys, market volatility, illiquidity, and possible discontinuation
of the Ecosystem.  Tokens and Node Deeds are not covered by investor‐
compensation schemes under Directive~97/9/EC or deposit‐guarantee schemes
under Directive~2014/49/EU.  The Company is not liable for losses arising
from such risks.

\vspace{0.5em}
\noindent\textbf{Liability limits.}
We disclaim all warranties—express or implied—regarding title,
merchantability, fitness for purpose, or technical quality of the Tokens
and licences.  We shall not be liable for any direct or indirect losses
arising from acquisition or use of Tokens, Node Deeds, or the Ecosystem, nor
for delays, modifications, or failures of any project component.

\vspace{0.5em}
\noindent\textbf{Forward-looking statements.}
This White Paper may contain forward-looking statements subject to factors
beyond our control.  Actual results may differ materially.  We assume no
liability for reliance on such statements.

\vspace{0.5em}
\noindent\textbf{Jurisdiction.}
This document and any relationship between \emph{you} and \emph{us} shall be
governed by the laws of Romania.  Distribution of this White Paper may be
restricted in certain jurisdictions; recipients must observe all applicable
laws and obtain any required advice before proceeding.

\section*{Acknowledgments}\label{ACKS}

We extend our sincere gratitude to the entire Ratio1.ai - including but not limited to \textbf{Traian Ispir, Serban Macrineanu, Mihai Constantinescu, Alberto Bastianello, Veaceslav Botezatu} (in no particular order) -  for their invaluable support in integrating advanced artificial intelligence with blockchain technology, revolutionizing the AI landscape by integrating cutting-edge artificial intelligence with blockchain technology. 
\\\\
\begin{quote}
\centering{\textbf{Ratio1.ai focuses on the fusion that facilitates the rapid development and deployment of AI applications through a decentralized, scalable, and user-owned-data platform.}}

\end{quote}

\bibliographystyle{unsrt}  
\bibliography{references}

\end{document}